\documentclass[11pt,twoside]{article}

\usepackage{asp2006}
\usepackage{fancyhdr,graphicx,caption,rotating,multirow,lscape}
\DeclareGraphicsExtensions{.eps,.eps.gz,.ps,.jpg,.tiff}

\begin{document}

\title{The colour of noise in SuperWASP data and the implications for finding
extra-solar planets}

\author{A. M. S. Smith,\altaffilmark{1} A. Collier Cameron,\altaffilmark{1} D. J. Christian,\altaffilmark{2} W. I. Clarkson,\altaffilmark{3} B. Enoch,\altaffilmark{4} A. Evans,\altaffilmark{5} C. A. Haswell,\altaffilmark{4} C. Hellier,\altaffilmark{5} K. Horne,\altaffilmark{1} J. Irwin,\altaffilmark{6} S. R. Kane,\altaffilmark{7} T. A. Lister,\altaffilmark{5} A. J. Norton,\altaffilmark{4} N. Parley,\altaffilmark{4} D. L. Pollacco,\altaffilmark{2} R. Ryans,\altaffilmark{2} I. Skillen,\altaffilmark{8} R. A. Street,\altaffilmark{2} A. H. M. J. Triaud,\altaffilmark{1} R. G. West,\altaffilmark{9} P. J. Wheatley,\altaffilmark{10} \& D. M. Wilson,\altaffilmark{5}}

\altaffiltext{1}{SUPA, School of Physics \& Astronomy, University of St. Andrews, North Haugh, 
St. Andrews, Fife, KY16 9SS, UK}
\altaffiltext{2}{Astrophysics Research Centre, Main Physics Building, School of Mathematics \&
Physics, Queen's University, Belfast, BT7 1NN, UK}

\altaffiltext{3}{Space Telescope Science Institute (STScI), 3700 San Martin Drive,
Baltimore, MD~21218, USA}

\altaffiltext{4}{Department of Physics \& Astronomy, The Open University, Milton Keynes, 
MK7 6AA, UK}

\altaffiltext{5}{Astrophysics Group, School of Chemistry \& Physics, Keele University, 
Staffs, ST5 5BG, UK}

\altaffiltext{6}{Institute of Astronomy, University of Cambridge, Madingley Road, 
Cambridge, CB3 0HA, UK}

\altaffiltext{7}{Department of Astronomy, University of Florida, 211 Bryant Space Science Center,
Gainesville, FL 32611-2055, USA}

\altaffiltext{8}{Isaac Newton Group of Telescopes, Apartado de correos 321,
E-38700 Santa Cruz de la Palma, Tenerife, Spain}

\altaffiltext{9}{Department of Physics \& Astronomy, University of Leicester, Leicester, 
LE1 7RH, UK}

\altaffiltext{10}{Department of Physics, University of Warwick, Coventry CV4 7AL, UK}

\begin{abstract}
A recent study demonstrated that there is significant covariance structure in
the noise on data from ground-based photometric surveys designed to detect
transiting extrasolar planets. Such correlation in the noise has often been
overlooked, especially when predicting the number of planets a particular survey
is likely to find. Indeed, the shortfall in the number of transiting extrasolar
planets discovered by such surveys seems to be explained by co-variance in the
noise. We analyse SuperWASP (Wide Angle Search for Planets) data and determine that
there is a significant amount of correlated systematic noise present. After
modelling the potential planet catch, we conclude that this noise places a
significant limit on the number
of planets that SuperWASP is likely to detect; and that the best way to boost the
signal-to-noise ratio and limit the impact of co-variant noise is to increase
the number of observed transits for each candidate transiting planet.
\end{abstract}

\section{Introduction}

Although a total of 14 transiting extra-solar planets have been discovered since
the first, HD209458b \citep{HD209}, the number of discoveries has not lived
up to early predictions \citep[e.g.][]{Horne}. This observed shortfall in the number of transiting planets detected has recently been explained by the presence of correlated
`red' noise in the data from ground-based transit surveys \citep{Pont}. Previous
forecasts of the planet `catch' from ground-based surveys assumed that the
noise in such data is entirely un-correlated or `white' in nature. Pont showed
that systematic red noise correlated on time-scales equivalent to a typical hot
Jupiter (HJ) transit ($\approx$ 2.5 hours) cannot be ignored and indeed tends to
be the dominant type of noise for bright stars.

\section{Observations}

SuperWASP-N is a wide-field transit survey instrument that started observing in 2004 at the Isaac Newton Group
on La Palma.
The observations used in this work are those made by SuperWASP-N in the 2004
season, when the instrument comprised an array of five lenses, each with a CCD recording a 7.8\hbox{$^\circ$} ~by 7.8\hbox{$^\circ$} ~field. Up to eight sets of fields were observed at a time, with measurements made about 7 - 8 minutes apart. Further details of
the SuperWASP project can be found in \cite{Pollacco}. Prior to searching the data for transits, the 
{\sc SysRem} algorithm of \cite{Tamuz} is applied to the data to reduce
systematic errors. Four components of red noise are removed with {\sc
SysRem}; it is thought that these are caused by variations in sky brightness,
vignetting, focus and residual secondary extinction \citep{CCmethods}.

\section{How much red noise?}

We follow closely the approach of \cite{PZQ06} in estimating the level of
correlated noise present in SuperWASP data. We compute the running average of
the data over the $n (=20)$ points contained in a transit-length time interval (2.5 hours). If the noise is purely random, the RMS scatter in the average of $n$ data points should be 
$\sigma_\mathrm{w}=\sigma / \sqrt{n}$, where $\sigma$ is the standard RMS of the whole lightcurve. If,
however, there is a systematic component in the noise, the RMS scatter of the
average of $n$ points, $\sigma_\mathrm{r}$, will be greater than this.

The quantities $\sigma$, $\sigma_\mathrm{r}$ and $\sigma_\mathrm{w}$ are calculated for each of the non-variable stars in one of the SuperWASP-N fields both prior to and after decorrelation with {\sc SysRem}. Fig. 1 shows that the {\sc SysRem} algorithm is highly effective at reducing the
levels of systematic noise present in the data, but that not all correlation in the noise is removed. If that
were the case, the $\sigma_\mathrm{r}$ curve would lie over the
$\sigma_\mathrm{w}$ curve. Instead, the $\sigma_\mathrm{r}$ curve lies
higher than $\sigma_\mathrm{w}$, and flattens out at about 3~mmag for bright
($V$ = 9.5) stars, indicating that systematic trends of this magnitude are present in
the data on a 2.5 hour timescale.

\begin{figure}[h]
\centering
\begin{minipage}[c]{.49\textwidth}
\includegraphics[angle=270,width=6.55cm]{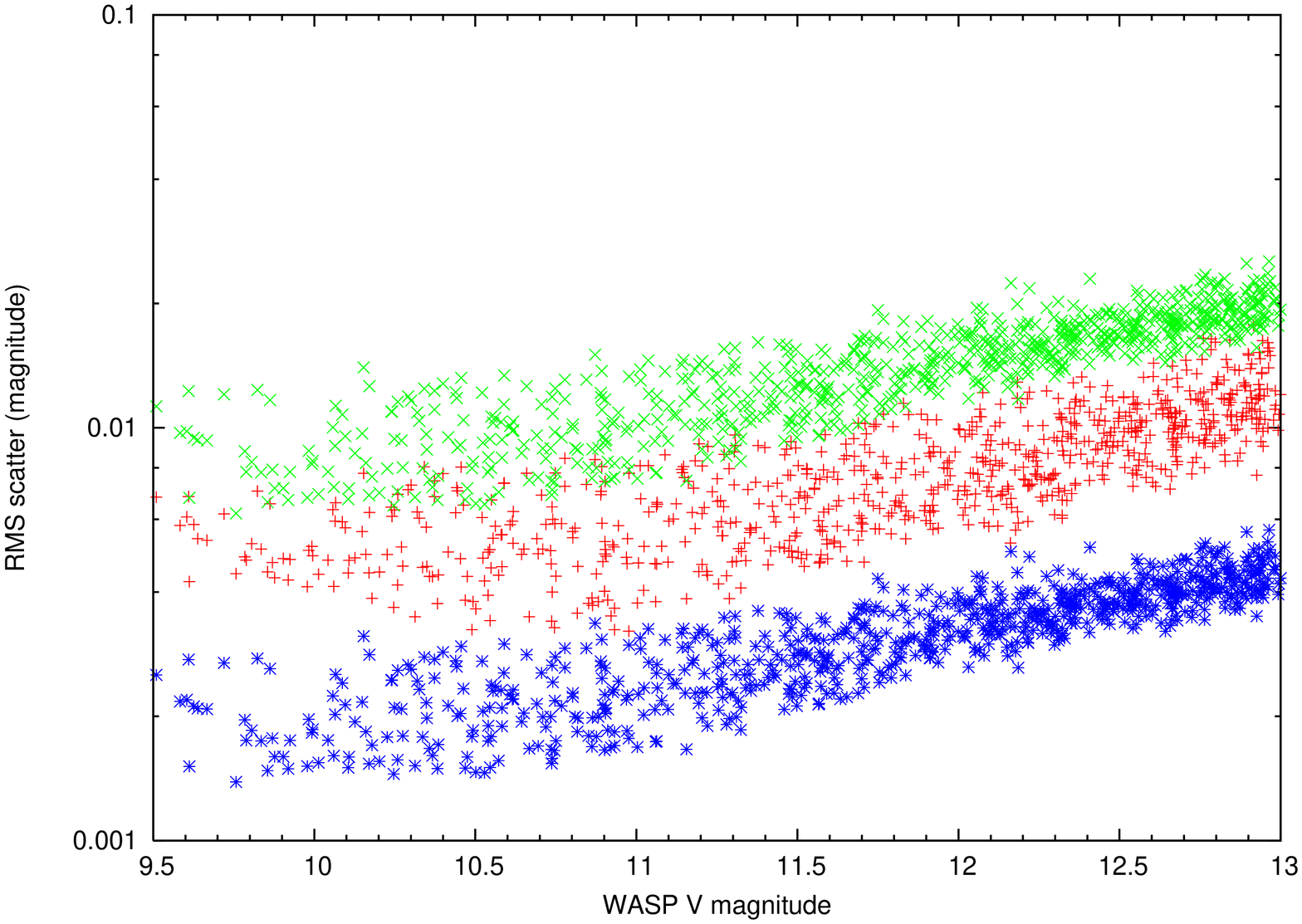}
\end{minipage}
\begin{minipage}[c]{.49\textwidth}
\includegraphics[angle=270,width=6.55cm]{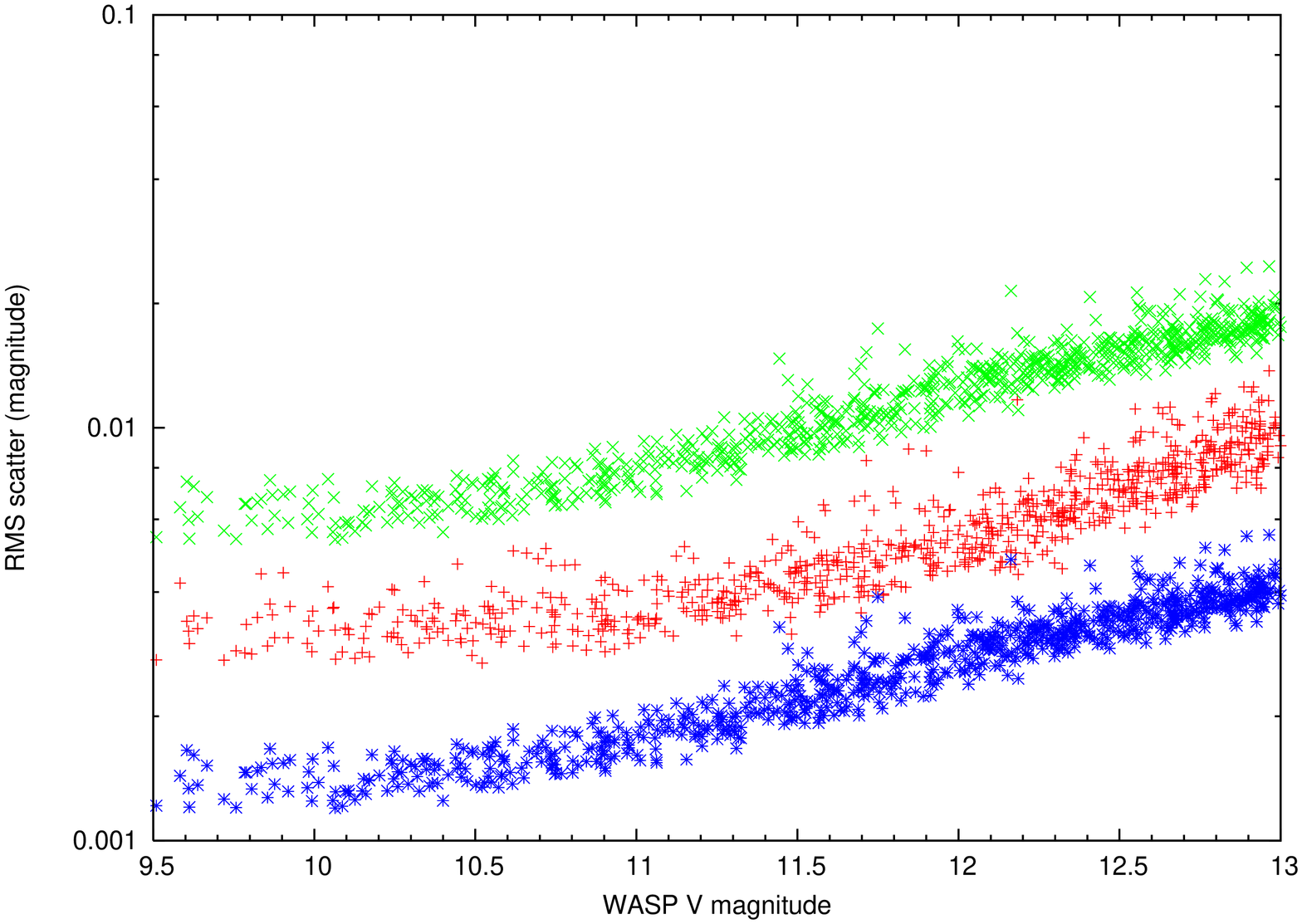}
\end{minipage}
\caption{RMS scatter versus magnitude for the non-variable stars in one field both prior to (left hand panel), and after
(right hand panel) decorrelation with {\sc SysRem}. The upper curve shows the RMS scatter of the
lightcurve of each object, $\sigma$. The middle curve shows the scatter,
$\sigma_\mathrm{r}$, after a 
moving average over a 2.5 hour time interval (20 data points) was calculated. The
lower curve shows the RMS scatter divided by $\sqrt{20}$, $\sigma_\mathrm{w}$.
}
\end{figure}

\section{Simulated planet catch}

We model the objects in the 20 fields observed by one of the SuperWASP cameras in the 2004
season, by using the Besan\c con model of the Galaxy
\citep{Robin} to generate a catalogue of stars with $9.5<V<13.0$ for each of
the fields. Planets are then assigned to stars that are of spectral class F, G or
K and luminosity class IV or V on the basis of their metallicity, using the
planet-metallicity relation of \cite{F+V05},
\begin{displaymath}
\mathcal{P}(\mathrm {planet}) =0.03\times10^{2.0[\mathrm{Fe/H}]},
\end{displaymath}
where $\mathcal{P}(\mathrm {planet})$ is the probability that a star of
metallicity [Fe/H] is host to a planet. This probability, along with a semi-major axis drawn from a uniform log distribution \citep{Smith}, is used to determine whether or not each star hosts a transiting planet. 151 $\pm 13$ of the 154,156 stars generated by the Besan\c con model are allocated a transiting planet. It is assumed,
for simplicity, that all planets have a radius, $R_{\mathrm {p}}$, equal to that of Jupiter.

In the regime where red noise dominates, the signal-to-noise ratio, $S_{\mathrm {red}}$, is 
given by
\begin{displaymath}
S_{\mathrm {red}}=\frac{\Delta m\sqrt{n_\mathrm{trans}}}{\sigma_\mathrm{r}(V)},
\end{displaymath}
where $n_\mathrm{trans}$ is the number of transits observed and $\Delta m$ is the transit depth, determined using equation (9) of \cite{T+S}. The red noise as a function of magnitude, $\sigma_\mathrm{r}(V)$, is determined by fitting a line to the middle curve of the right hand panel of Fig. 1. Using SuperWASP observation times enables us to calculate $n_\mathrm{trans}$ and hence $S_{\mathrm {red}}$ \citep{Smith}. 

The value of $S_{\mathrm {red}}$ that one chooses as a threshold for planetary detection is a compromise between allowing
too many false positive detections and rejecting large numbers of genuine transiting planets; in this work we use the (perhaps slightly conservative) requirement that $S_{\mathrm {red}} \ge 10$ for a planet to be `detected'.

The signal-to-noise ratio for each simulated planet is calculated
for each of three different observing baselines; the results are plotted in Fig. 2. Also shown is the detection efficiency as a function of period for the
requirement that 2, 4, and 6 transits are observed. Increasing the number of transits required for a detection causes the detection efficiency
to fall dramatically at most 
periods. If one requires a larger number (6 or more) of transits for
detection, then the detection efficiency is much lower, except for several pathological
periods where the detection fraction 
is such that finding planets with that period is particularly favourable.
Increasing the number of observing nights has the effect of increasing the detection
efficiency at nearly all periods (Fig. 2).

\begin{figure}[ht!]
\centering
\begin{minipage}[c]{.49\textwidth}
\includegraphics[angle=270,width=6.55cm]{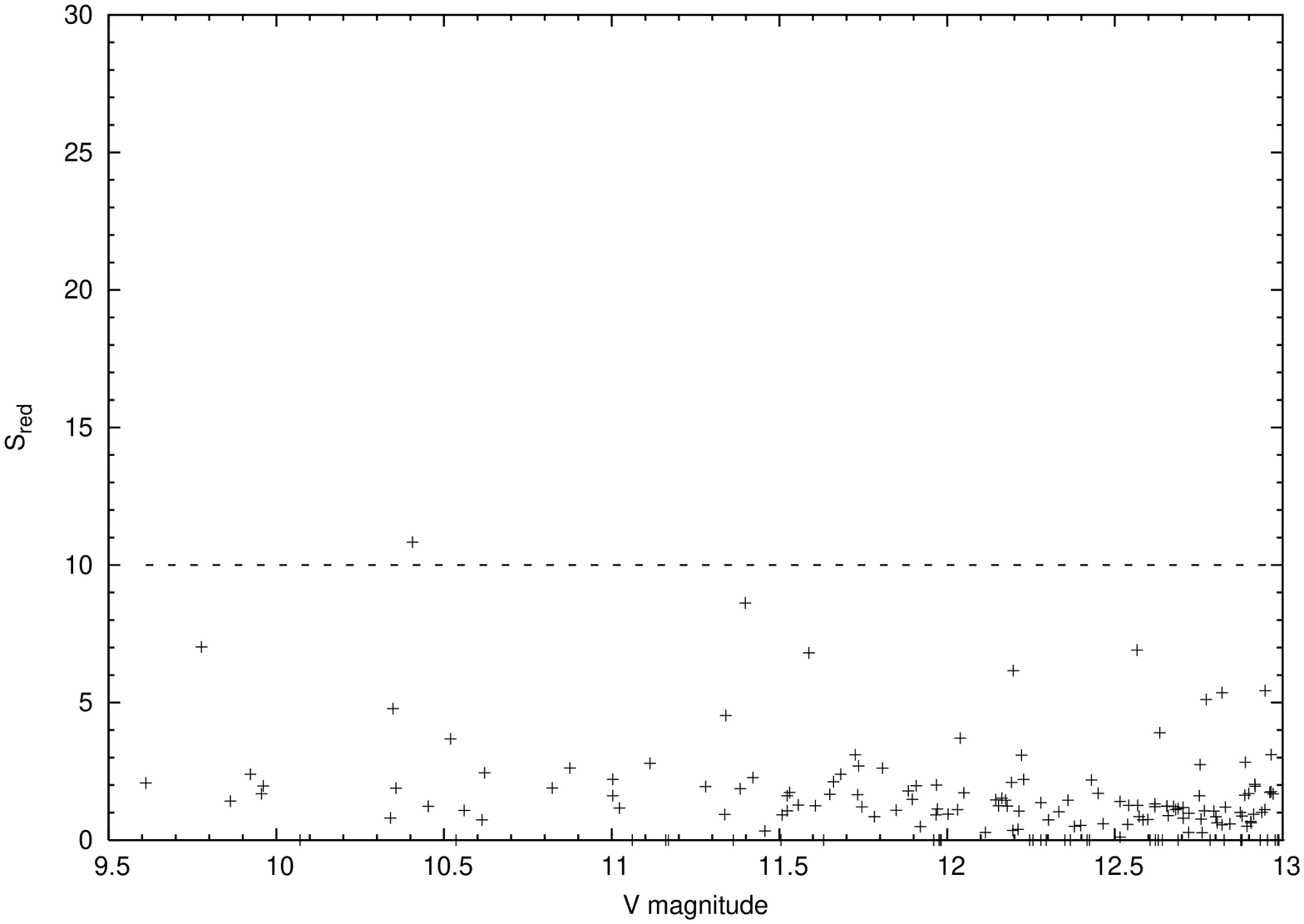}
\end{minipage}
\begin{minipage}[c]{.49\textwidth}
\includegraphics[angle=270,width=6.55cm]{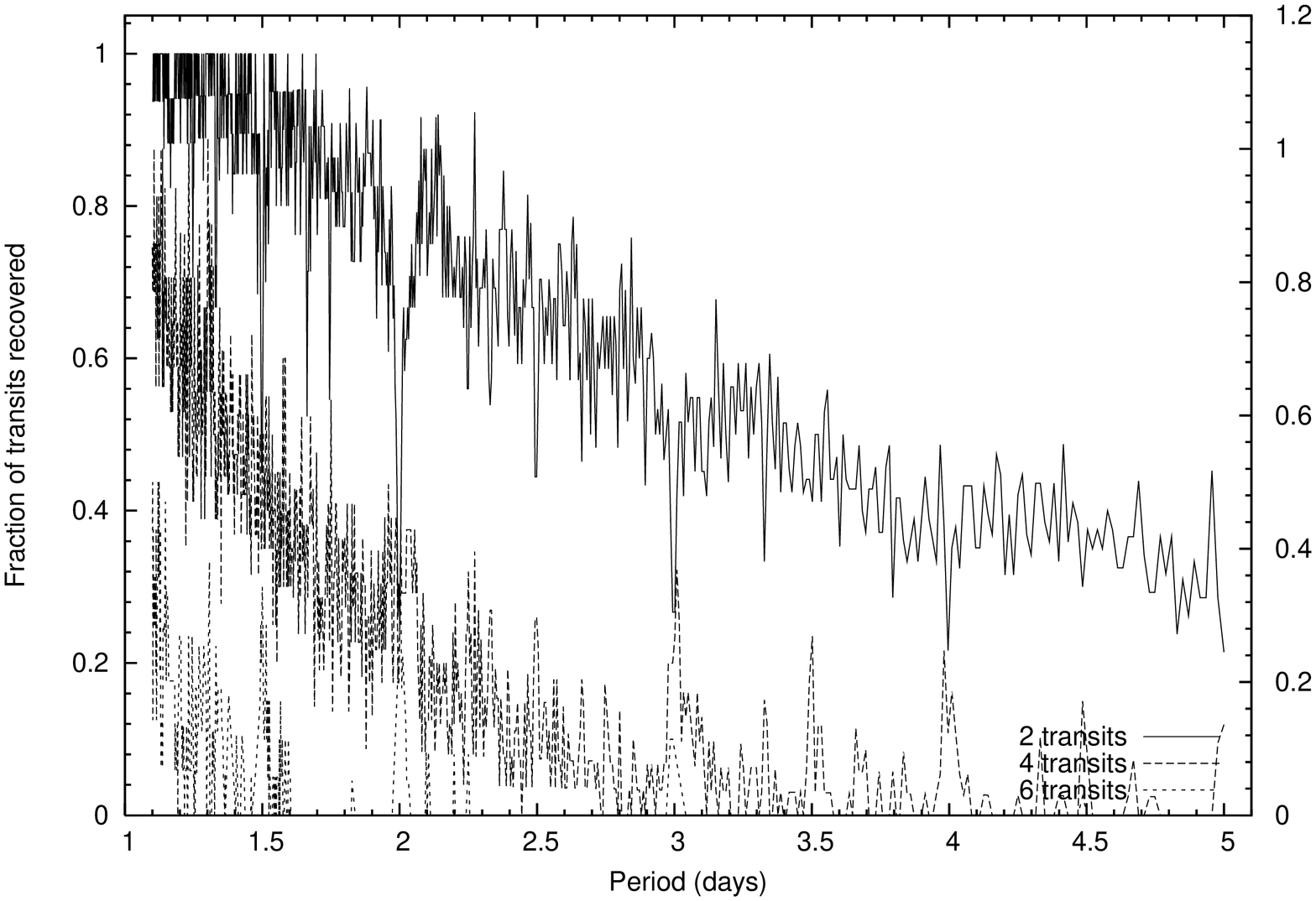}
\end{minipage}
\begin{minipage}[c]{.49\textwidth}
\includegraphics[angle=270,width=6.55cm]{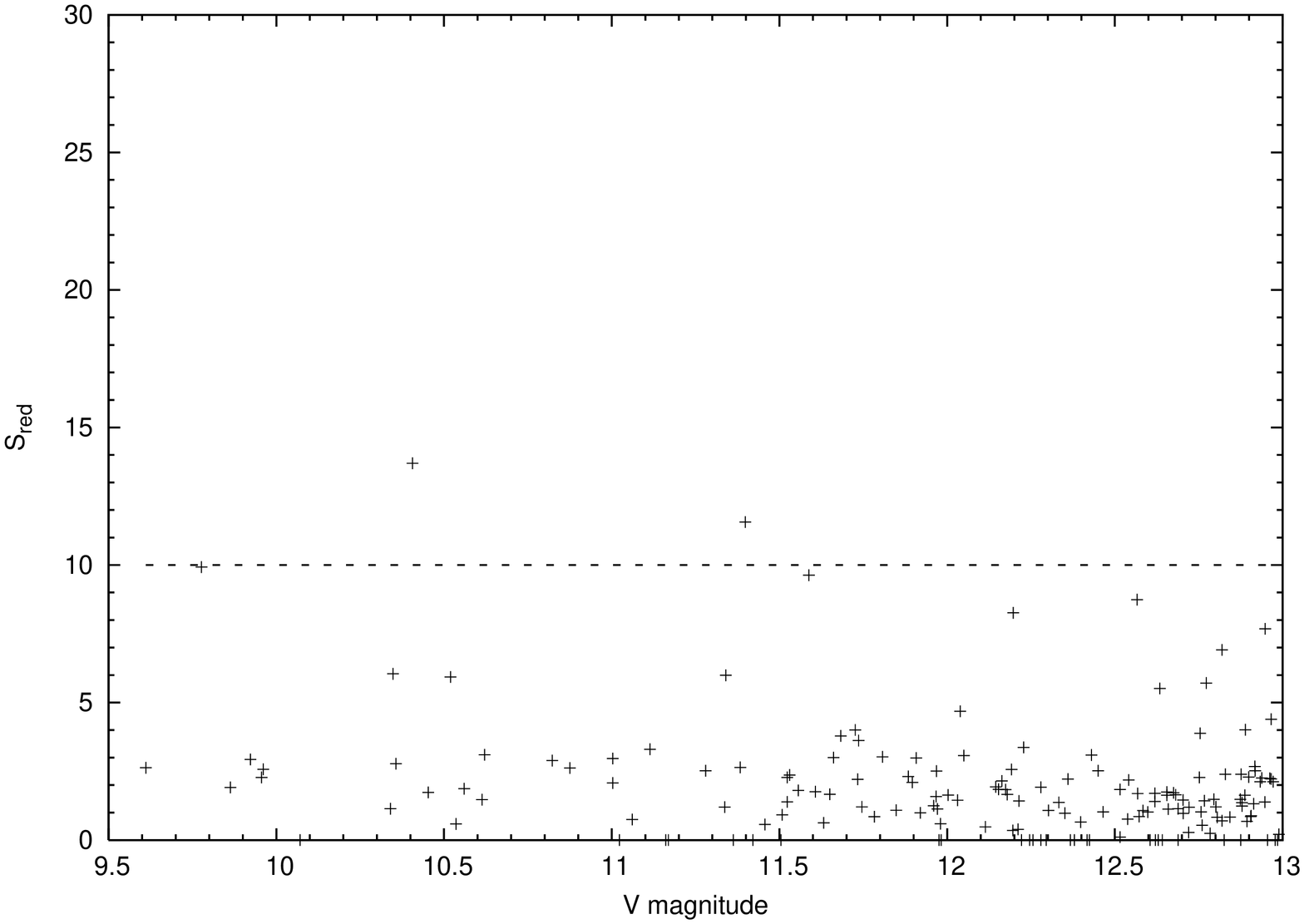}
\end{minipage}
\begin{minipage}[c]{.49\textwidth}
\includegraphics[angle=270,width=6.55cm]{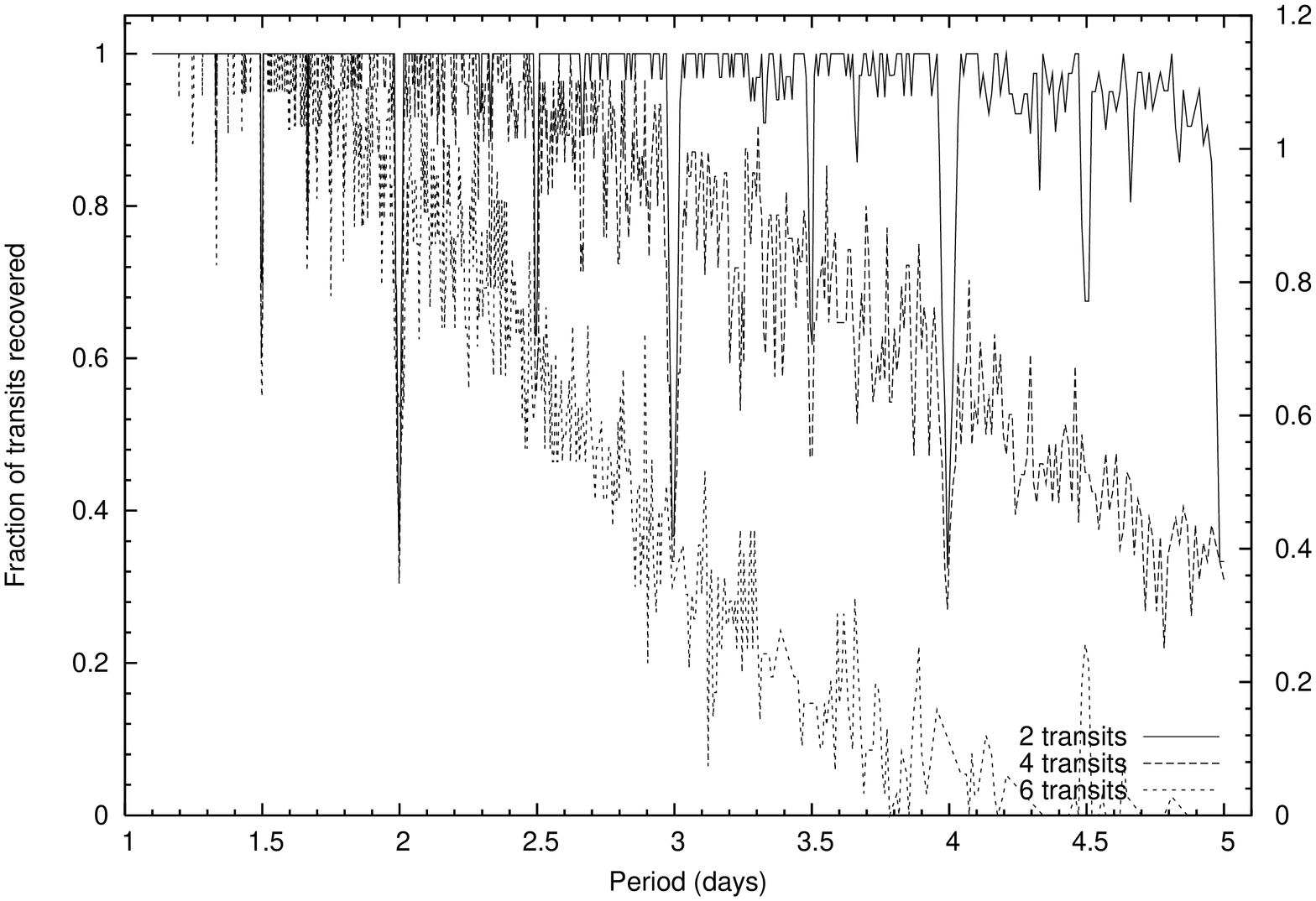}
\end{minipage}
\begin{minipage}[c]{.49\textwidth}
\includegraphics[angle=270,width=6.55cm]{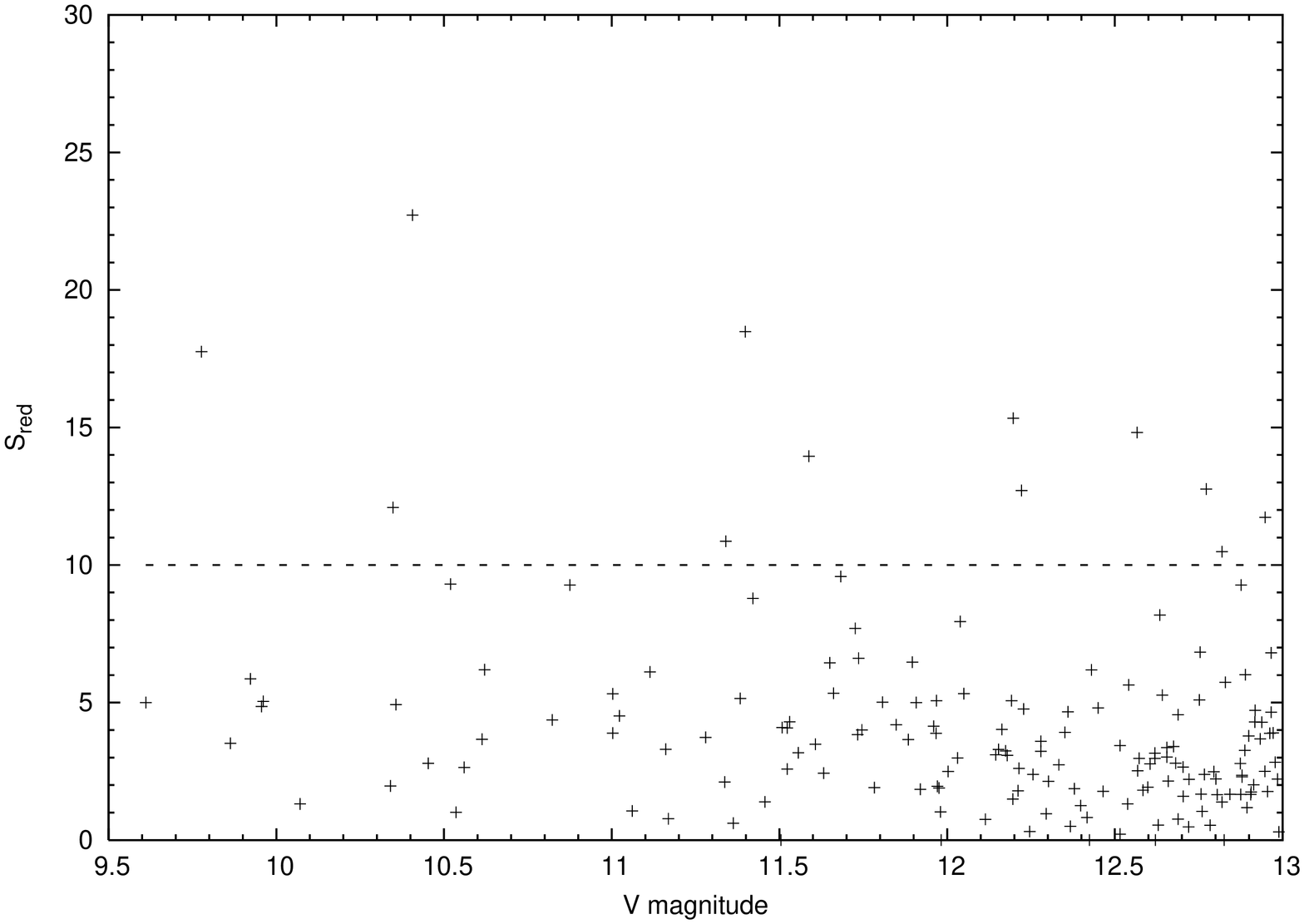}
\end{minipage}
\begin{minipage}[c]{.49\textwidth}
\includegraphics[angle=270,width=6.55cm]{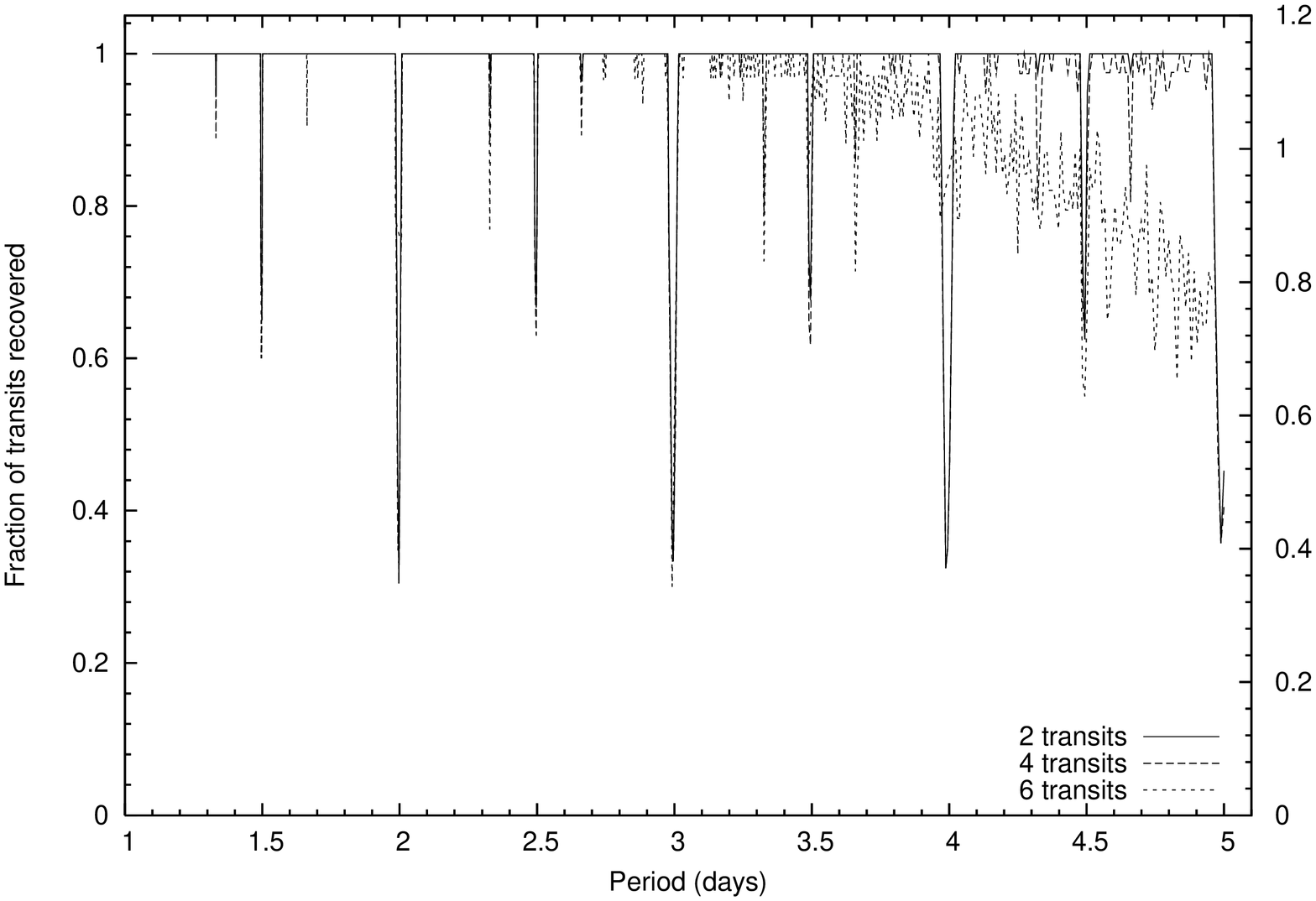}
\end{minipage}

\caption{Left hand panels: Signal-to-noise ratio versus magnitude for 151 simulated
transiting extra-solar planets, with the $S_{\mathrm {red}}=10$ threshold
indicated. Top-to-bottom: 51, 80, 130 nights of observations result in 1, 2, and 12
`detections' respectively. Right hand panels:
the corresponding transit detection efficiency as a function of period. In each case the solid,
upper curve is for the requirement that
at least 2 transits are observed for a detection; the dashed, middle curve for 4
transits; and the dotted, lower curve for 6.
}
\end{figure}

A notable property of
many of the current SuperWASP transit candidates \citep[e.g.][]{Christian06} is their large value of $n_{\mathrm{trans}}$. Also, many of the candidate periods coincide with the narrow pathological period ranges
where there is a much greater chance of detecting a large number of transits.
The effects of red noise can be reduced by increasing the signal-to-noise of the
data by requiring that a larger number of transits $(\approx 10)$ are observed.
In order to have a reasonable chance of observing this many transits, especially
for planets with periods that are not pathologically favourable, longer
time-base observations are required.

A further simulation of the 20 SuperWASP-N fields, this time using the observing
baseline of each field, yields a total of 3.72 $\pm$ 1.60 detected planets (see
Smith et al. 2006 for the complete results of this simulation).
The detection rate was found to increase linearly with observing baseline
(Fig.~3).

\begin{figure}[!ht]
\centering
\includegraphics[angle=270,width=6.55cm]{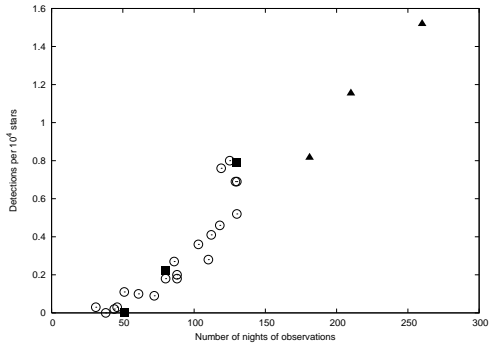}
\caption{Detection rate of transiting extra-solar planets versus length of
observing baseline. The squares correspond to all 20 fields modelled using the same
observing baseline, while the triangles represent an additional 130 nights of
observations in a second season. The 20 fields, modelled using their own
observing baselines, are represented by the open
circles.}
\end{figure}

Given that SuperWASP-N observed with five cameras in 2004, our predicted planet
yield can be scaled up to 18.6 $\pm$ 8.0 planets, which we compare to the
number of planets discovered by SuperWASP-N from the 2004 data set. Two new
transiting systems, WASP-1 and WASP-2, were discovered \citep[\citeauthor{Street} \citeyear{Street};][]{CCplanet} and a third, known system, XO-1 \citep{XO-1} was
also detected.

There are several reasons for the apparent discrepancy between the number of
predicted planets and the number detected. Probably most significant is that
only the best candidates brighter than $V = 12$ were selected for the initial
radial velocity follow up. Analysis of our model reveals that $\approx 25$ per
cent of our predicted detections lie in the magnitude range $12 < V < 13$, and
further undetected planets may lie among the remaining brighter candidates.
Additionally, the distribution of semi-major axis used in our model \citep{Smith}
may tend to slightly over-predict the number of very short-period planets, to which the
transit method is most sensitive.

\section{Conclusions}

We conclude that there is a significant component of systematic, red noise
present in data from SuperWASP-N. The {\sc SysRem} algorithm of
Tamuz et al. (2005)
appears highly effective at reducing the level of red noise, but fails to
eliminate it entirely. The remaining 3~mmag of red noise present in the data on transit duration time-scales has a significant impact on planet detection.

Modelling the objects in the fields of one SuperWASP-N camera reveals that in order to  improve the $S_{\mathrm {red}}$, and thus increase the number of detectable planets, a greater number of transits must be observed in the data set of a particular object. This requires observations  to be made over a longer time period.

On the basis of our predicted transit detection rates, the SuperWASP consortium have decided to continue observing for a further season all the fields that were monitored during 2004. We expect that this will enhance greatly the number of planetary transit events detected at non-pathological periods.

\acknowledgements

The WASP Consortium consists of representatives from the Universities of Cambridge (Wide
Field Astronomy Unit), Keele, Leicester, The Open University, Queens University Belfast and
St Andrews, along with the Isaac Newton Group (La Palma) and the Instituto de
Astrof\'{i}sica de
Canarias (Tenerife). The SuperWASP cameras were constructed and are operated with
funds made available from the Consortium Universities and PPARC. AMSS wishes to acknowledge the
financial support of a UK PPARC studentship.

\end{document}